\documentclass
[aps,pra,amsfonts,amssymb,twocolumn,amsmath,preprintnumbers,nofootinbib,floatfix,
showpacs,superscriptaddress]{revtex4-1}%
\usepackage[dvips]{graphics}
\usepackage{graphicx}
\usepackage{bm}
\usepackage{amsmath}
\usepackage{amsfonts}
\usepackage{amssymb}%
\providecommand{\U}[1]{\protect\rule{.1in}{.1in}}

\begin{document}

\title{Rashba torque beyond the Boltzmann regime}
\author{Cong Xiao}
\affiliation{Department of Physics, The University of Texas at Austin, Austin, Texas 78712, USA}

\author{Qian Niu}
\affiliation{Department of Physics, The University of Texas at Austin, Austin, Texas 78712, USA}
\affiliation{School of Physics, ICQM and CICQM, Peking University, Beijing 100871, China}
\begin{abstract}
We study spin torques induced by Rashba spin-orbit coupling in two-dimensional
ferromagnets under the good-metal condition $\epsilon_{F}\tau/\hbar\gg1$
($\epsilon_{F}$ the Fermi energy, $\tau$ the electron lifetime) by employing
the Kubo formula. We find that, in the presence of spin-dependent disorder the
Rashba torque changes greatly as the system evolves out of the weak disorder
limit where $\hbar/\tau$ is much smaller than any intrinsic energy scale
characterizing the multiband structure. The antidamping-like component of
Rashba torque can be comparable to and larger than the field-like one out of the
weak disorder limit. The semiclassical Boltzmann theory produces the same
results as microscopic linear response calculations only in the weak disorder
limit. Our analysis indicates that rich behaviors of various nonequilibrium
phenomena beyond the Boltzmann theory may also be present even when
$\epsilon_{F}\tau/\hbar\gg1$ in multiband systems where $\epsilon_{F}$ is not
the unique intrinsic energy scale.
\end{abstract}
\maketitle


\section{Introduction}

In describing nonequilibrium phenomena of conduction electrons in solids, the
widely-employed semiclassical Boltzmann theory \cite{Ziman} is valid only if
the band structure is well-defined, i.e., the disorder broadening $\hbar/\tau$
($\tau$ the electron lifetime) of bands is much smaller than any intrinsic
energy scale characterizing the multiband structure
\cite{Sinitsyn2008,Xiao2017SOT,Luttinger}. This Boltzmann regime is often
called the weak disorder limit \cite{Sinitsyn2008,Xiao2017SOT,Luttinger} or
weak scattering limit \cite{Nunner2007}. In simple systems where the Fermi
energy $\epsilon_{F}$ is the unique intrinsic energy scale characterizing the
conduction band such as the conventional parabolic band \cite{Lee1985} and
linear Dirac band \cite{Sinitsyn}, the Boltzmann regime is practically
equivalent to the good-metal limit $\epsilon_{F}\tau/\hbar\gg1$.

In multiband systems, some intrinsic energy scales characterizing the
multiband structure of conduction bands may not be large compared to
$\hbar/\tau$, nonequilibrium phenomena then exhibit behaviors beyond the
Boltzmann regime even in the good-metal limit $\epsilon_{F}\tau/\hbar\gg1$.
This is the possible case in systems with spin-orbit coupling, such as the
Rashba spin-orbit coupling which widely exists in inversion-asymmetric
structures \cite{Rashba}. In a Rashba system with both subbands partially
occupied, the band splitting $\Delta_{k}$ due to the Rashba coupling (and
exchange coupling in ferromagnetic Rashba systems) provides an intrinsic
energy-scale \cite{Liu2011}. Because the Rashba coupling is weak in many cases
\cite{Rashba}, the competition between $\Delta_{k}$ and $\hbar/\tau$ can lead
to rich behaviors of nonequilibrium phenomena beyond the Boltzmann regime.

In the present paper we reveal such behaviors of Rashba torque
\cite{Li2015,Titov2015,Manchon2012} in the 2D ferromagnetic Rashba model with
both subbands partially occupied. Under the good-metal condition $\epsilon
_{F}\tau/\hbar\gg1$ there are still two different limits
\cite{Liu2011,Culcer2013}: the weak disorder limit $\Delta_{k}\gg\hbar/\tau$
and the opposite limit $\Delta_{k}\ll\hbar/\tau$. This latter limit is often
called the diffusive limit \cite{Burkov2004}, because in this limit the spin
relaxation time is much larger than the momentum relaxation time and thus when
considering spin dynamics the motion of electron is diffusive.

The Rashba torque arises from the s-d coupling between the spin of Rashba
electrons and the local magnetization: an applied electric field induces a
nonequilibrium spin density of conduction electrons via the Rashba spin-orbit
coupling, this spin density then exerts a torque on the local magnetization.
The Rashba torque possesses two components: a field-like torque odd in the
magnetization direction $\mathbf{\hat{M}}$ and an antidamping-like torque even
in $\mathbf{\hat{M}}$ \cite{Li2015}. Although there have been plenty of
researches on the Rashba torque in the weak disorder limit
\cite{Li2015,Titov2015,Xiao2017SOT} or diffusive limit \cite{Manchon2012},
some basic characters have not been revealed, such as the possibility that the
antidamping-like component becomes larger than the field-like one and the
evolution of Rashba torque from the weak disorder limit to the diffusive
limit. Besides, most papers on the Rashba torque employing the Boltzmann
theory or other phenomenological treatment
\cite{Zhang2008,Duine2012,Lee2015,Kim2012} did not point out which limit (weak
disorder limit or diffusive limit) their theories work in.

We consider the 2D Rashba model with a perpendicular magnetization
\cite{Titov2015} in order to avoid the complexity induced by the in-plane
anisotropy. This anisotropy is important for the angular dependence of Rashba
torque \cite{Lee2015}, but is not the interest here. By employing the Kubo
formula under the non-crossing approximation, we find that both the field-like
and antidamping-like Rashba torques change considerably from the weak disorder
limit to the diffusive limit, provided that the disorder is not completely
spin-independent. Especially, the antidamping-like torque, which is much
smaller than the field-like one in the weak disorder limit, becomes comparable to
and even larger than the field-like one out of the weak disorder limit.

In the discussion part of this paper we also address other nonequilibrium
phenomena such as the anomalous Hall effect \cite{Nunner2007}, spin Hall
effect \cite{Sinova2015} and Edelstein effect \cite{Edelstein1990} in 2D
Rashba systems. In all these cases, we show that the semiclassical Boltzmann
theory is consistent with microscopic linear response calculations only in the
weak disorder limit rather than in the whole regime of the good-metal limit.

The present paper is organized as follows. General formulations are presented
in Sec. II, whereas calculation results are given and analyzed in Sec. III.
Section IV makes some discussions and concludes the paper.

\section{Formulation}

\subsection{Model}

The 2D ferromagnetic Rashba Hamiltonian is%
\begin{equation}
\hat{H}_{0}=\frac{\hbar^{2}\mathbf{k}^{2}}{2m}+\alpha_{R}\mathbf{\hat{\sigma}%
}\cdot\left(  \mathbf{k}\times\mathbf{\hat{z}}\right)  -J_{ex}\mathbf{\hat
{\sigma}\cdot\hat{M}}, \label{model}%
\end{equation}
where $m$ is the effective mass of conduction electron, $\mathbf{k}=k\left(
\cos\phi,\sin\phi\right)  $ the 2D wavevector, $\mathbf{\hat{\sigma}}=\left(
\hat{\sigma}_{x},\hat{\sigma}_{y},\hat{\sigma}_{z}\right)  $ are the Pauli
matrices, $\alpha_{R}$ is the Rashba parameter, $J_{ex}$ the exchange
coupling, the direction of the local magnetization is chosen to be
$\mathbf{\hat{M}=\hat{z}}$ for the isotropic model. We only consider the case
$\epsilon_{F}>J_{ex}$\textbf{,} i.e., both Rashba bands partially occupied
(more accurately, we demand $\epsilon_{F}-J_{ex}\gg\hbar/\tau$). For any
energy $\epsilon>J_{ex}$ there are two iso-energy rings corresponding to the
two Rashba bands $\eta=\pm$: $k_{\eta}^{2}\left(  \epsilon\right)  =\frac
{2m}{\hbar^{2}}\left(  \epsilon-\eta\Delta_{\eta}\left(  \epsilon\right)
\right)  $ where $\epsilon_{R}=m\left(  \frac{\alpha_{R}}{\hbar}\right)  ^{2}$
and $\Delta_{\eta}\left(  \epsilon\right)  \equiv\Delta_{k_{\eta}\left(
\epsilon\right)  }=\sqrt{J_{ex}^{2}+\alpha_{R}^{2}k_{\eta}^{2}\left(
\epsilon\right)  }=\sqrt{\epsilon_{R}^{2}+J_{ex}^{2}+2\epsilon_{R}\epsilon
}-\eta\epsilon_{R}$. The density of states in $\eta$ band is $D_{\eta}\left(
\epsilon\right)  =D_{0}\frac{\Delta_{\eta}\left(  \epsilon\right)  }%
{\Delta_{\eta}\left(  \epsilon\right)  +\eta\epsilon_{R}}$ with $D_{0}%
=\frac{m}{2\pi\hbar^{2}}$.

The short-range (pointlike) disorder can be classified according to the spin
structure of scattering potential as \cite{Yang2011}: class A $\hat{V}%
=V_{A}\hat{\sigma}_{0}$, class B $\hat{V}=V_{B}\hat{\sigma}_{z}$, class C
$\hat{V}=V_{c}\hat{\sigma}_{\pm}/\sqrt{2}$. Here $\hat{\sigma}_{\pm}=$
$\hat{\sigma}_{x}\pm i\hat{\sigma}_{y}$, $\hat{\sigma}_{0}$ is the identity
matrix in spin space. It was shown that in the weak disorder limit the
contribution from class A disorder to the anomalous Hall effect is quite
different from that of classes B and C disorder, even with opposite sign
\cite{Yang2011}. While the contributions from class B and C disorder are
similar \cite{Yang2011}. Thus our calculation only takes into account class A
and class B disorder. We will show that the spin structure of short-range
disorder strongly affects the behavior of Rashba torque when the system
evolves from the weak disorder limit to the diffusive limit under the
good-metal condition. For this purpose it is sufficient to assume Gaussian
disorder \cite{Titov2015,Titov2017,Manchon2012}.

\subsection{Kubo-Streda formalism}

In the linear response analysis, the average value of an observable $A$
(Hermitian operator $\hat{A}$, which can represent a vector, scalar, etc) in
the presence of a dc uniform weak electric field $\mathbf{E}$ and disorder is
generally given by \cite{Streda1977} $A=Tr\left\langle \hat{\rho}^{0}\hat
{A}^{0}\right\rangle _{c}+Tr\left\langle \left(  \delta^{\mathbf{E}}\hat{\rho
}\right)  \hat{A}^{0}\right\rangle _{c}+Tr\left\langle \hat{\rho}^{0}%
\delta^{\mathbf{E}}\hat{A}\right\rangle _{c}$, where $\hat{\rho}=\hat{\rho
}^{0}+\delta^{\mathbf{E}}\hat{\rho}$ is the total density matrix,
$\left\langle ..\right\rangle _{c}$ denotes the average over disorder
configurations and $Tr$ the trace over relevant degrees of freedom. $\hat
{\rho}^{0}$ and $\hat{A}^{0}$ are operators in equilibrium, $\delta
^{\mathbf{E}}\hat{\rho}$ and $\delta^{\mathbf{E}}\hat{A}$ represent the
out-of-equilibrium change of operators linear in $\mathbf{E}$.\ The last term
of $A$ is relevant usually in thermal related effects, such as the heat
current response to an electric field \cite{Streda1977}. Here we do not
consider thermal related effects and focus on the case $\delta^{\mathbf{E}%
}\hat{A}=0$. Then%
\begin{equation}
\delta A=Tr\left\langle \hat{A}^{0}\left(  \delta^{\mathbf{E}}\hat{\rho
}\right)  \right\rangle _{c}, \label{general}%
\end{equation}
with $\delta A=A-A_{0}$ and $A_{0}\equiv Tr\left\langle \hat{\rho}^{0}\hat
{A}^{0}\right\rangle _{c}$.

The linear response (\ref{general}) in the single-particle picture with only
elastic electron-impurity scattering can be found by the Kubo-Streda formula
\cite{Bruno2001,Ebert2015} for the correlation function between $\hat{A}$ and
the electric current operator $\mathbf{\hat{\jmath}}=e\mathbf{\hat{v}}$. For
instance, at low-temperature limit the Kubo-Streda formula for the
electric-field induced nonequilibrium spin density $\delta S_{\alpha}=$
$\chi_{\alpha\beta}E_{\beta}$ reads \cite{Ebert2015,Titov2017} $\chi
_{\alpha\beta}=\chi_{\alpha\beta}^{I\left(  a\right)  }+\chi_{\alpha\beta
}^{I\left(  b\right)  }+\chi_{\alpha\beta}^{II}$, where
\begin{equation}
\chi_{\alpha\beta}^{I\left(  a\right)  }=\frac{\hbar}{2\pi}Tr\left\langle
\hat{S}_{\alpha}\hat{G}^{R}\left(  \epsilon_{F}\right)  \hat{\jmath}_{\beta
}\hat{G}^{A}\left(  \epsilon_{F}\right)  \right\rangle _{c}, \label{surface-a}%
\end{equation}%
\begin{equation}
\chi_{\alpha\beta}^{I\left(  b\right)  }=-\frac{\hbar}{2\pi}\operatorname{Re}%
Tr\left\langle \hat{S}_{\alpha}\hat{G}^{R}\left(  \epsilon_{F}\right)
\hat{\jmath}_{\beta}\hat{G}^{R}\left(  \epsilon_{F}\right)  \right\rangle
_{c}, \label{surface-b}%
\end{equation}%
\begin{gather}
\chi_{\alpha\beta}^{II}=\frac{\hbar}{2\pi}\operatorname{Re}\int d\eta
f^{0}\left(  \epsilon\right) \label{sea}\\
\times Tr\left\langle \hat{S}_{\alpha}\hat{G}^{R}\left(  \epsilon\right)
\hat{\jmath}_{\beta}\frac{d\hat{G}^{R}\left(  \epsilon\right)  }{d\epsilon
}-\hat{S}_{\alpha}\frac{d\hat{G}^{R}\left(  \epsilon\right)  }{d\epsilon}%
\hat{\jmath}_{\beta}\hat{G}^{R}\left(  \epsilon\right)  \right\rangle
_{c}.\nonumber
\end{gather}
Here $\alpha,\beta=x,y$, $\hat{S}_{\alpha}$ is the $\alpha$-component of spin
operator, $\hat{G}^{R/A}\left(  \epsilon\right)  =\left(  \epsilon-\hat{H}\pm
i0^{+}\right)  ^{-1}$ is the retarded/advanced Green's function operator with
$\hat{H}=\hat{H}_{0}+\hat{V}$, $f^{0}$\ is the Fermi distribution function.

\subsection{Formal expressions for Rashba torque}

In the case of class A (B) disorder, the imaginary part of the retarded Born
self-energy $\Sigma^{R}=\sum_{\mathbf{k}^{\prime}}V_{\mathbf{kk}^{\prime}%
}G_{\mathbf{k}^{\prime}}^{0,R}V_{\mathbf{k}^{\prime}\mathbf{k}}$\ is diagonal
in spin space and inversely proportional to the electron lifetime
$\tau=\left(  \frac{2\pi}{\hbar}n_{im}^{A\left(  B\right)  }V_{A\left(
B\right)  }^{2}D_{0}\right)  ^{-1}$ with $n_{im}^{A\left(  B\right)  }$ the
density of class A (B) disorder. The dressed retarded Green's function is then
\cite{Nunner2007} $G_{\mathbf{k}}^{R}=\sum_{i}G_{i\mathbf{k}}^{R}\hat{\sigma
}_{i}$ with $G_{0\mathbf{k}}^{R}=\frac{1}{2}\sum_{\eta}G_{\eta k}^{R}$,
$G_{x\mathbf{k}}^{R}=\frac{\sin\theta\sin\phi}{2}\sum_{\eta}\eta G_{\eta
k}^{R}$, $G_{y\mathbf{k}}^{R}=-\frac{\sin\theta\cos\phi}{2}\sum_{\eta}\eta
G_{\eta k}^{R}$, $G_{z\mathbf{k}}^{R}=-\frac{\cos\theta}{2}\sum_{\eta}\eta
G_{\eta k}^{R}$ and $G_{\eta k}^{R}\left(  \epsilon_{F}\right)  =\left(
\epsilon_{F}-\epsilon_{k}^{\eta}+\frac{i\hbar}{2\tau}\right)  ^{-1}$. Here
$\cos\theta=J_{ex}/\Delta_{k}$.

Under the good-metal condition, $\chi_{\alpha\beta}^{II}$\ and $\chi
_{\alpha\beta}^{I\left(  b\right)  }$ are approximated by their disorder-free
parts \cite{Nunner2007,noteSea} which are zero in the considered case. Thus we
get $\chi_{\alpha\beta}=\chi_{\alpha\beta}^{I\left(  a\right)  }$, which is
calculated by bubble with ladder vertex corrections \cite{Liu2011,Burkov2004}
under the non-crossing approximation: $\chi_{\alpha y}=\frac{\hbar e}{2\pi
}\frac{\hbar}{2}\sum_{\mathbf{k}}tr\left[  \sigma_{\alpha}G_{\mathbf{k}}%
^{R}\left(  \epsilon_{F}\right)  \Upsilon_{y}G_{\mathbf{k}}^{A}\left(
\epsilon_{F}\right)  \right]  $, with $\Upsilon_{y}$ the dressed velocity
vertex and $tr$ the trace in spin space.

In this section we only consider the presence of class A or class B disorder
alone. Assuming $\Upsilon_{y}=a\frac{\hbar k_{y}}{m}+b\sigma_{x}+c\sigma_{y}$
with $a,b,c$ real numbers, we get
\begin{align}
\chi_{yy}  &  =-\frac{\hbar e}{\pi}\frac{\hbar}{2}\frac{2\pi\tau D_{0}}{\hbar
}\left(  ibI_{2}-cI_{1}\right)  ,\nonumber\\
\chi_{xy}  &  =\frac{\hbar e}{\pi}\frac{\hbar}{2}\frac{2\pi\tau D_{0}}{\hbar
}\left(  aI_{3}+bI_{1}+ciI_{2}\right)  ,
\end{align}
where $I_{3}=-\alpha_{R}/\hbar$ and
\begin{align}
\text{\ }I_{1}  &  =\sum_{\eta}\left[  \sin^{2}\theta_{\eta}+\frac{1+\cos
^{2}\theta_{\eta}}{1+\left(  2\Delta_{\eta}\left(  \epsilon_{F}\right)
\frac{\tau}{\hbar}\right)  ^{2}}\right]  \frac{D_{\eta}\left(  \epsilon
_{F}\right)  }{4D_{0}},\nonumber\\
I_{2}  &  =-i\frac{\tau}{\hbar}\sum_{\eta}\frac{J_{ex}}{1+\left(
2\Delta_{\eta}\left(  \epsilon_{F}\right)  \frac{\tau}{\hbar}\right)  ^{2}%
}\frac{D_{\eta}\left(  \epsilon_{F}\right)  }{D_{0}}. \label{I-1}%
\end{align}
Here $\cos\theta_{\eta}=\frac{J_{ex}}{\Delta_{\eta}\left(  \epsilon
_{F}\right)  }$, $\sin\theta_{\eta}=\frac{\alpha_{R}k_{\eta}\left(
\epsilon_{F}\right)  }{\Delta_{\eta}\left(  \epsilon_{F}\right)  }$. We note
that $I_{1}$ and $I_{2}$ depend on the parameter $2\Delta_{\eta}\left(
\epsilon_{F}\right)  \frac{\tau}{\hbar}$ which measures the competition
between the intrinsic multiband splitting and the disorder-induced band
broadening. In the weak disorder limit $2\Delta_{\eta}\left(  \epsilon
_{F}\right)  \gg\hbar/\tau$, the multiband structure is well-defined, the
picture of intraband and interband processes is clear \cite{Kovalev2010} thus
the semiclassical Boltzmann theory born in the band-eigenstate representation
works well \cite{Xiao2017SOT}. However, when $2\Delta_{\eta}\left(
\epsilon_{F}\right)  \lesssim\hbar/\tau$, the intrinsic multiband splitting of
two Rashba bands is overwhelmed by the disorder broadening. This case,
although still in the good-metal limit (the longitudinal electrical
conductivity is still large due to $\epsilon_{F}\tau/\hbar\gg1$), cannot be
well treated by the Boltzmann theory.

For later convenience, here we work out
\begin{equation}
I_{1}\simeq\frac{\epsilon_{R}\epsilon_{F}}{J_{ex}^{2}+2\epsilon_{R}%
\epsilon_{F}},I_{2}\simeq\frac{-i\hbar}{2\tau}\frac{J_{ex}}{J_{ex}%
^{2}+2\epsilon_{R}\epsilon_{F}} \label{I Boltzmann}%
\end{equation}
in the weak disorder limit $\hbar/\tau\ll\Delta_{\eta}\left(  \epsilon
_{F}\right)  ,\epsilon_{F}$, and
\begin{equation}
I_{1}\simeq1-2\left(  1+\frac{J_{ex}^{2}}{\Delta_{0}^{2}}\right)  \left(
\bar{\Delta}\frac{\tau}{\hbar}\right)  ^{2},\text{ }I_{2}\simeq-i\frac{2\tau
}{\hbar}J_{ex} \label{I diffusive}%
\end{equation}
in the diffusive limit $\Delta_{\eta}\left(  \epsilon_{F}\right)  \ll
\hbar/\tau\ll\epsilon_{F}$. Here $\bar{\Delta}=\frac{1}{2}\sum_{\eta}%
\Delta_{\eta}\left(  \epsilon_{F}\right)  $ and the diffusive-limit condition
also implies that $\epsilon_{R},J_{ex},\sqrt{\epsilon_{R}\epsilon_{F}}\ll
\hbar/\tau\ll\epsilon_{F}$.

\section{Results}

\subsection{Class A disorder}

In this case the dressed velocity vertex is given by $\Upsilon_{y}=v_{y}%
+\frac{\hbar}{2\pi\tau D_{0}}\sum_{\mathbf{k}^{\prime}}G_{\mathbf{k}^{\prime}%
}^{R}\left(  \epsilon_{F}\right)  \Upsilon_{y}G_{\mathbf{k}^{\prime}}%
^{A}\left(  \epsilon_{F}\right)  $, yielding $a=1$ and $b=c=0$. Then%
\begin{equation}
\chi_{yy}=0,\text{ }\chi_{xy}=-e\alpha_{R}D_{0}\tau. \label{SOT A}%
\end{equation}
They remain the same forms from the weak disorder limit to the diffusive
limit. This result is the same as that obtained in previous diagrammatic
calculation \cite{Titov2017} and consistent with the weak-disorder-limit
result in a recently formulated semiclassical Boltzmann theory
\cite{Xiao2017SOT}.

Besides, under the same approximations, the anomalous Hall effect vanishes
$\sigma_{xy}=0$ and the longitudinal conductivity takes the Drude form
$\sigma_{yy}=n_{e}e^{2}\tau/m$, where $n_{e}=2D_{0}\left(  \epsilon
_{F}+\epsilon_{R}\right)  $ is the carrier density. This two quantities also
remain the same forms from the weak disorder limit to the diffusive limit.
These results indicate that the scalar short-range disorder is very special in
a Rashba system in the case of both bands partially occupied under the
non-crossing approximation.

\subsection{Class B disorder}

In this case $\Upsilon_{y}=v_{y}+\frac{\hbar}{2\pi\tau D_{0}}\sum
_{\mathbf{k}^{\prime}}\sigma_{z}G_{\mathbf{k}^{\prime}}^{R}\Upsilon
_{y}G_{\mathbf{k}^{\prime}}^{A}\sigma_{z}$ yields $a=1$, $b=2\frac{\alpha_{R}%
}{\hbar}\frac{1+I_{1}}{\left(  1+I_{1}\right)  ^{2}+\left(  iI_{2}\right)
^{2}}$ and $c=\frac{iI_{2}}{1+I_{1}}b$. Then%
\begin{align}
\chi_{yy}  &  =-e\alpha_{R}D_{0}\tau\frac{2iI_{2}}{\left(  1+I_{1}\right)
^{2}+\left(  iI_{2}\right)  ^{2}},\nonumber\\
\chi_{xy}  &  =-e\alpha_{R}D_{0}\tau\frac{1-I_{1}^{2}-\left(  iI_{2}\right)
^{2}}{\left(  1+I_{1}\right)  ^{2}+\left(  iI_{2}\right)  ^{2}}.
\label{SOT B gen}%
\end{align}
Both\ depend on the competition between the intrinsic band-splitting and
disorder broadening.

In the weak disorder limit, by Eq. (\ref{I Boltzmann}), we get%
\begin{align}
\chi_{yy}  &  =-\hbar e\alpha_{R}D_{0}\frac{J_{ex}\left(  J_{ex}^{2}%
+2\epsilon_{R}\epsilon_{F}\right)  }{\left(  J_{ex}^{2}+3\epsilon_{R}%
\epsilon_{F}\right)  ^{2}},\nonumber\\
\chi_{xy}  &  =-e\alpha_{R}D_{0}\tau\frac{J_{ex}^{2}+\epsilon_{R}\epsilon_{F}%
}{J_{ex}^{2}+3\epsilon_{R}\epsilon_{F}}, \label{SOT B Boltzmann}%
\end{align}
which confirms the result obtained by a recent Boltzmann theory
\cite{Xiao2017SOT}.

In the diffusive limit, by Eq. (\ref{I diffusive}), we have%
\begin{equation}
\chi_{xy}=-e\alpha_{R}D_{0}\tau\left(  \bar{\Delta}\frac{\tau}{\hbar}\right)
^{2},\text{ }\chi_{yy}=-e\alpha_{R}D_{0}\tau\frac{\tau J_{ex}}{\hbar},
\label{SOT B diffusive}%
\end{equation}
which strongly depend on the electron lifetime $\tau$ and are totally beyond
the semiclassical Boltzmann theory. The ratio between the longitudinal and
transverse nonequilibrium spin densities is very large (here we assume the
exchange coupling is not too small $J_{ex}\gtrsim\sqrt{\epsilon_{R}%
\epsilon_{F}}$):
\begin{equation}
\frac{\chi_{yy}}{\chi_{xy}}=\frac{J_{ex}}{\bar{\Delta}}\frac{\hbar}%
{\bar{\Delta}\tau}\gg1.
\end{equation}
\begin{figure}[ptbh]
\includegraphics[width=0.35\textwidth]{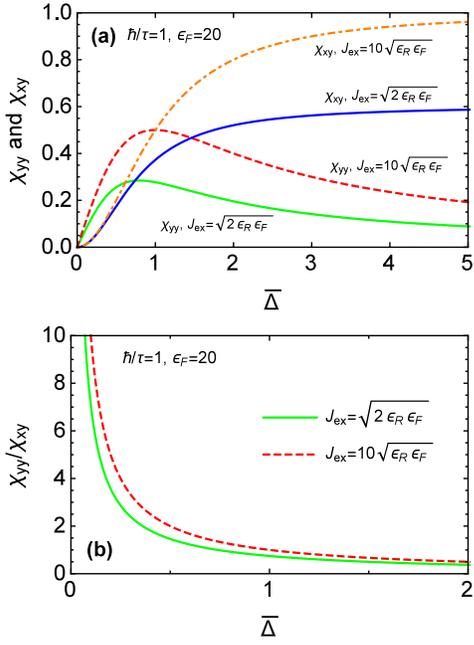} \caption{The evolution
of (a) $\chi_{yy}$ and $\chi_{xy}$ as well as (b) their ratio $\chi_{yy}%
/\chi_{xy}$ from the weak disorder limit ($\bar{\Delta}\tau/\hbar\gg1$) to the
diffusive limit ($\bar{\Delta}\tau/\hbar\ll1$) in the case of class B
disorder, $\bar{\Delta}=\frac{1}{2}\sum_{\eta}\Delta_{\eta}\left(
\epsilon_{F}\right)  $. The ratio $\chi_{yy}/\chi_{xy}$ represents the
relative strength of antidamping-like and field-like components of the Rashba
torque. Here and in Fig. 2 we assume $\epsilon_{R}\ll\epsilon_{F}$ and thus
use the parameter $\bar{\Delta}\tau/\hbar$ to control the evolution from the
weak disorder limit to the diffusive limit. We choose $\hbar/\tau=1$,
$\epsilon_{F}=20$ in plotting the curves for the case of $J_{ex}%
=\sqrt{2\epsilon_{R}\epsilon_{F}}$ and $J_{ex}=10\sqrt{\epsilon_{R}%
\epsilon_{F}}$. In (a) $\chi_{yy}$ and $\chi_{xy}$ are measured in the units
of $-e\alpha_{R}D_{0}\tau$.}%
\label{fig1}%
\end{figure}

The evolution of $\chi_{yy}$ and $\chi_{xy}$ as well as their ratio from the
weak disorder limit to the diffusive limit is plotted in Fig. 1 where we set
$\hbar/\tau=1$ as the unit of energy and $\epsilon_{F}=20$. In plotting Fig. 1
(and Fig. 2) we assume $\epsilon_{R}\ll\epsilon_{F}$ and thus can use the
parameter $\bar{\Delta}\tau/\hbar$ to approximately control the evolution from
the weak disorder limit $\bar{\Delta}\tau/\hbar\gg1$ to the diffusive limit
$\bar{\Delta}\tau/\hbar\ll1$. To simplify the analysis, we choose
$J_{ex}=\sqrt{2\epsilon_{R}\epsilon_{F}}$ and $J_{ex}=10\sqrt{\epsilon
_{R}\epsilon_{F}}$ in plotting the curves. In Fig. 1(a) $\chi_{yy}$ and
$\chi_{xy}$ are measured in the units of $-e\alpha_{R}D_{0}\tau$. Figure. 1
shows that both $\chi_{yy}$ (corresponds to an antidamping-like Rashba torque
in the direction $\mathbf{\hat{M}}\times\left[  \left(  \mathbf{\hat{z}}%
\times\mathbf{E}\right)  \times\mathbf{\hat{M}}\right]  $) and $\chi_{xy}$
(corresponds to a field-like Rashba torque in the direction $\mathbf{\hat{M}%
}\times\left(  \mathbf{\hat{z}}\times\mathbf{E}\right)  $) change greatly from
the weak disorder limit to the diffusive limit, and $\chi_{yy}/\chi_{xy}$
rapidly increases when the system evolves towards the diffusive limit. The
non-monotonicity of $\chi_{yy}$ in Fig. 1(a) is just what can be expected from
the two limiting values of $\chi_{yy}$ in Eqs. (\ref{SOT B Boltzmann}) and
(\ref{SOT B diffusive}) in the case of $J_{ex}\propto\sqrt{\epsilon
_{R}\epsilon_{F}}$.

For comparison, here we also present the values of electrical conductivities
in the diffusive limit: $\sigma_{yy}=\frac{e^{2}}{\pi\hbar}\frac{\tau}{\hbar
}\left(  \epsilon_{F}+\epsilon_{R}\right)  $, $\sigma_{xy}=\frac{e^{2}}%
{\pi\hbar}\frac{2\tau J_{ex}}{\hbar}\frac{\epsilon_{R}\tau}{\hbar}$, and
$\sigma_{xy}/\sigma_{yy}\simeq\frac{2\tau J_{ex}}{\hbar}\frac{\epsilon_{R}%
}{\epsilon_{F}}\ll1$. Unlike $\sigma_{yy}$ whose leading contribution under
the good-metal condition is always proportional to $\epsilon_{F}\tau/\hbar$
(not shown), $\chi_{xy}$ is not proportional to $\epsilon_{F}$ even in the
weak disorder limit. Thus as the system evolves from the weak disorder limit
to the diffusive limit, while $\sigma_{yy}$ remains large, $\chi_{xy}$ may
become much smaller and may not remain dominant over $\chi_{yy}$. As a result,
in complex multiband systems with multiple intrinsic energy scales under the
good-metal condition $\epsilon_{F}\tau/\hbar\gg1$, the relaxation time
approximation (or lifetime approximation) of the Boltzmann equation can be
used as a first approximation for analyzing $\sigma_{yy}$ even out of the
Boltzmann regime (weak disorder limit), producing a result proportional to
$\tau$. But this is not the case for analyzing $\chi_{xy}$ which is of higher
order in $\tau$ out of the Boltzmann regime unless in rare special cases such
as that considered in Sec. III. A. This important difference between
$\sigma_{yy}$ and $\chi_{xy}$ explains why the Boltzmann analysis of the
longitudinal charge (and thermoelectric) transport is usually qualitatively
valid even out of the weak disorder limit in complex multiband systems
\cite{Ziman}, and also indicates that this kind of \textquotedblleft
extended\textquotedblright\ validity of the semiclassical Boltzmann theory may
not be possible for other nonequilibrium phenomena which exhibit rich
behaviors out of the weak disorder limit. This point will be further discussed
in Sec. IV. Also, we can conclude that, the relaxation time approximation of 
the Boltzmann equation as a first approximation for the analysis of 
nonequilibrium phenomena is likely to be qualitatively valid only in the 
weak disorder limit (except for the longitudinal electric transport).

\subsection{Competition between classes A and B}

In the presence of both class A and class B impurities, we assume
$\left\langle V_{A}V_{B}\right\rangle _{c}=0$ following Ref.
\onlinecite{Yang2011} and expect that interference effects between class A and
B scattering do not qualitatively alter the result in this subsection. Here we
only give the main results, calculation details are present in Appendix
A.\begin{figure}[ptbh]
\includegraphics[width=0.4\textwidth]{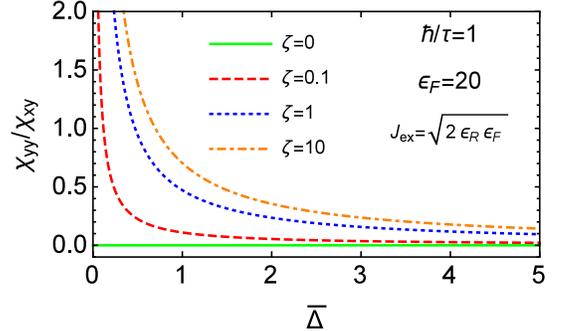} \caption{
$\chi_{yy}/\chi_{xy}$ for fixed values of $\varsigma$ in the presence of both
class A and B disorder from the weak disorder limit to the diffusive limit.
Here $\hbar/\tau=1$, $\epsilon_{F}=20$ and $J_{ex}=\sqrt{2\epsilon_{R}%
\epsilon_{F}}$.}%
\label{fig4}%
\end{figure}

Due to $\sum_{\eta}\frac{D_{\eta}}{\Delta_{\eta}}=0$, the electron lifetime is
still simple $\tau=\tau_{A}/\left(  1+\zeta\right)  =\left(  \tau_{A}%
^{-1}+\tau_{B}^{-1}\right)  ^{-1}$, where $1/\tau_{A\left(  B\right)  }=2\pi
n_{im}^{A\left(  B\right)  }V_{A\left(  B\right)  }^{2}D_{0}/\hbar$, and
$\zeta=\tau_{A}/\tau_{B}$ represents the relative weight of the two scattering
classes. The $\zeta$-dependence of $\chi_{yy}/\chi_{xy}$ from the weak
disorder limit to the diffusive limit is shown in Fig. 2. As $\zeta$ increases
from zero, the curve of $\chi_{yy}$ is shifted upward from the class A
dominated regime due to the increasing contribution from class B scattering.

According to the results in this section, even under the good-metal condition
$\epsilon_{F}\tau/\hbar\gg1$ the Rashba torque exhibits rich behaviors, and
the spin structure of disorder strongly affects the behavior of Rashba torque
in the whole good-metal regime.

\section{Discussion and conclusion}

We start this discussion section by pointing out that some previous papers
(on the Rashba torque) that did not emphasize which limit (diffusive limit or weak disorder limit)
they work in are in fact within the weak disorder limit. References
\onlinecite{Zhang2008,Duine2012,Lee2015} that employed the semiclassical
Boltzmann theory should be considered in the weak disorder limit according to
the analysis in the present paper. Reference \onlinecite{Kim2012} assumed the
exchange coupling is the largest energy scale affecting the conduction
electron spin dynamics in metallic Rashba ferromagnets, thus $2\Delta_{\eta
}\left(  \epsilon_{F}\right)  \simeq2J_{ex}\gg\hbar/\tau$ and this paper also
works in the weak disorder limit.

We note that rich behaviors from the weak disorder limit to the diffusive
limit can be found in various nonequilibrium phenomena in 2D Rashba systems
with both subbands partially occupied provided that the scattering is not
completely spin-independent. One can thus consider the spin Hall and anomalous
Hall effects in the case of magnetic impurities \cite{Wang2007,Ren2008}, as
well as the spin Hall and Edelstein effects in the presence of spin-orbit
scattering off scalar pointlike impurities \cite{Culcer2013,Raimondi2012}. In
all these cases, one can verify that the semiclassical Boltzmann theory only
applies in the weak disorder limit.

For example, in the presence of spin-orbit scattering off scalar disorder, the
semiclassical Boltzmann calculation \cite{note-SO scattering} yields zero spin
Hall conductivity in the first order of Rashba parameter, same as microscopic
linear response calculations in the weak disorder limit \cite{Culcer2013}.
Whereas in the diffusive limit $\bar{\Delta}\tau/\hbar\ll1$ nonzero spin Hall
effect due to the Elliot-Yafet spin relaxation caused by spin-orbit scattering
\cite{Raimondi2012} cannot be reproduced in the Boltzmann theory. Also,
applying the semiclassical Boltzmann framework proposed in Ref.
\onlinecite{Xiao2017SOT} to the spin Hall effect in the presence of isotropic
pointlike magnetic impurities just yields the weak-disorder-limit value
$\frac{-e}{7\pi}$ of the spin Hall conductivity obtained in Kubo diagrammatic
calculations (Eq. (21) in Ref. \onlinecite{Wang2007}), while the spin Hall
conductivity vanishes as $\left(  \bar{\Delta}\tau/\hbar\right)  ^{2}$ in the
diffusive limit.

Some final remarks are in order.

First, the diagrammatic calculation of the anomalous Hall effect in model
(\ref{model}) in the weak disorder limit has been improved by going beyond the
non-crossing approximation \cite{Ado}. How this additional contribution
changes from the weak disorder limit to the diffusive limit is left for future
work. Quantitative correction from this contribution to our results about the
Rashba torque is expected, but the qualitative aspect of the results in the
present paper is not expected to change.

Second, if one considers finite-range or long-range disorder, other fine
details besides the spin structure of disorder potentials should also be
carefully treated.

Third, in the ferromagnetic Rashba model with a perpendicular magnetization
where $J_{ex}<\epsilon_{R}$, there is a \textquotedblleft
window\textquotedblright\ \cite{Onoda} around the avoided band-anticrossing
point. The height of this window also provides an intrinsic energy scale
characterizing the conduction band. Rich behaviors of the anomalous Hall
effect beyond the Boltzmann regime when the Fermi energy is located in this
window have been deeply investigated \cite{Onoda}. One can verify that, in the
weak disorder limit (represented by $2J_{ex}\gg\hbar/\tau$ in this case), the
semiclassical Boltzmann theory \cite{Sinitsyn2008} produces the same results
as microscopic linear response theories \cite{Onoda,Nunner2007}. However, rich
behaviors of the anomalous Hall effect out of the weak disorder limit (called
superclean case in Ref. \onlinecite{Onoda}) obtained in the Keldysh approach
are beyond the scope of the Boltzmann theory. The band-anticrossing regime is
relevant in the case of strong Rashba coupling that is possible in
heavy-elements-related inversion-asymmetric structures. The behavior of Rashba
torque in this energy regime is also left for future work.

In summary, we have studied the Rashba torque in 2D Rashba ferromagnets under
the good-metal condition $\epsilon_{F}\tau/\hbar\gg1$ by employing the
Kubo-Streda formalism in the non-crossing approximation. It was shown that the
widely-used semiclassical Boltzmann theory produces the same results as the
Kubo formula only in the weak disorder limit. As the system evolves from the
weak disorder limit to the diffusive limit, both the antidamping-like and
field-like components of Rashba torque remain sensitive to the spin structure
of disorder. The magnitude of the antidamping-like component can be comparable to
and larger than the field-like one out of the weak disorder limit provided
that the short-range disorder is not completely spin-independent. We expect
these findings are helpful also in understanding spin-orbit torques in 2D
anti-ferromagnetic Rashba model \cite{Zelezny2017}.

The rich behaviors of nonequilibrium phenomena, like those in Rashba systems,
can also be expected in other multiband systems where the Fermi energy is not
the unique intrinsic energy scale characterizing the band structure of
conduction bands.

\begin{acknowledgments}
We acknowledge useful discussions with H. Chen. This work is supported by NBRPC (Grant No. 2013CB921900), DOE (DE-FG03-02ER45958, Division of Materials Science and Engineering), NSF (EFMA-1641101) and Welch Foundation (F-1255).
The formulation in Sec. II is supported by the DOE grant.
\end{acknowledgments}

\appendix

\section{Calculation details in the presence of both class A and B disorder}

The dressed velocity vertex is given by $\Upsilon_{y}=v_{y}+n_{im}^{A}%
V_{A}^{2}\sum_{\mathbf{k}^{\prime}}\left[  G_{\mathbf{k}^{\prime}}^{R}%
\Upsilon_{y}G_{\mathbf{k}^{\prime}}^{A}+\zeta\sigma_{z}G_{\mathbf{k}^{\prime}%
}^{R}\Upsilon_{y}G_{\mathbf{k}^{\prime}}^{A}\sigma_{z}\right]  $. Then $a=1$,
$b=\frac{\alpha_{R}}{\hbar}+\left(  1-\zeta\right)  bI_{1}^{\prime}+\left(
1-\zeta\right)  icI_{2}^{\prime}+\left(  1-\zeta\right)  I_{3}^{\prime}$ and
$c=\left(  1-\zeta\right)  cI_{1}^{\prime}-\left(  1-\zeta\right)
ibI_{2}^{\prime}$, where $I_{i}^{\prime}=\frac{1}{1+\zeta}I_{i},i=1,2,3$. Here
$I_{i}$ take the forms $I_{3}=\frac{\hbar}{2\pi\tau D_{0}}\sum_{\mathbf{k}%
}\frac{\hbar k_{y}}{m}2\operatorname{Re}G_{0}^{R}G_{x}^{A}$ and $I_{1\left(
2\right)  }=\frac{\hbar}{2\pi\tau D_{0}}\sum_{\mathbf{k}}\left(  G_{0}%
^{R}G_{0\left(  z\right)  }^{A}-G_{z}^{R}G_{z\left(  0\right)  }^{A}\right)
$. They are given by Eq. (\ref{I-1}) with $\tau^{-1}=\left(  1+\zeta\right)
\tau_{A}^{-1}$. 

Thus%
\[
c=-\frac{\frac{1-\zeta}{1+\zeta}ibI_{2}}{1-\frac{1-\zeta}{1+\zeta}I_{1}}%
=\frac{\alpha_{R}}{\hbar}\frac{-\frac{1-\zeta}{1+\zeta}\frac{2\zeta}{1+\zeta
}iI_{2}}{\left(  1-\frac{1-\zeta}{1+\zeta}I_{1}\right)  ^{2}+\left(
\frac{1-\zeta}{1+\zeta}\right)  ^{2}\left(  iI_{2}\right)  ^{2}},
\]
then $\chi_{yy}=\hbar eD_{0}\tau c\frac{1+\zeta}{1-\zeta}$ and
\[
\frac{\chi_{yy}}{\chi_{xy}}=\frac{2\zeta}{1+\zeta}\frac{iI_{2}}{\left(
1-\frac{1-\zeta}{1+\zeta}I_{1}\right)  \left(  1-I_{1}\right)  +\frac{1-\zeta
}{1+\zeta}\left(  iI_{2}\right)  ^{2}}.
\]

\end{document}